\documentclass[aps,prl,amsmath,amsfonts,amssymb,twocolumn,showpacs]{revtex4}
\usepackage[latin1]{inputenc}
\usepackage{graphicx}
\usepackage[english]{babel}

\usepackage{amsfonts}
\usepackage{amsmath}
\usepackage{amssymb}

\newcommand{\beq}{\begin{equation}}
\newcommand{\eeq}{\end{equation}}

\newcommand{\ket}[1]{|#1\rangle}
\newcommand{\bra}[1]{\langle #1|}

\makeatletter
\@ifundefined{textcolor}{}
{%
 \definecolor{BLACK}{gray}{0}
 \definecolor{WHITE}{gray}{1}
 \definecolor{RED}{rgb}{1,0,0}
 \definecolor{GREEN}{rgb}{0,1,0}
 \definecolor{BLUE}{rgb}{0,0,1}
 \definecolor{CYAN}{cmyk}{1,0,0,0}
 \definecolor{MAGENTA}{cmyk}{0,1,0,0}
 \definecolor{YELLOW}{cmyk}{0,0,1,0}
 }

\begin{document}

\title{The most elementary heat to work conversion: \\
work extraction from a completely passive state by a single photon}

\author{D. Valente
$^{1}$
}
\email{daniel@fisica.ufmt.br}

\author{F. Brito
$^{2}$
}

\author{T. Werlang
$^{1}$
}
\email{thiago\_werlang@fisica.ufmt.br}

\affiliation{
$^{1}$ 
Instituto de F\'isica, Universidade Federal de Mato Grosso, CEP 78060-900, Cuiab\'a, MT, Brazil
}

\affiliation{
$^{2}$ 
Instituto de F\'isica de S\~ao Carlos, Universidade de S\~ao Paulo, C.P. 369, 13560-970, S\~ao Carlos, SP, Brazil
}

\begin{abstract}
The concepts of work and heat in the quantum domain, as well as their interconversion principles, are still an open debate.
We have found theoretical evidence that a single photon packet is capable of extracting work from a single two-level system (TLS).
More importantly, this effect is found for a photon as it is spontaneously emitted, therefore carrying only heat away from the emitter.
This is the most elementary process in which heat is converted into work, requiring not more than two off-resonance atoms for that purpose.
From a more practical point of view it is found that, surprisingly, work can be extracted from a TLS in a completely passive initial state.
The process is cyclic in the sense that the TLS initial and final states are equal.
The state of the TLS remains passive throughout the interaction time with the single-photon packet.
The physical meaning of the work performed by the TLS is found to be a dynamical change in the color of the photon during the reemission process.
All our predictions are, in principle, measurable within state-of-the-art experiments.
\end{abstract}


\maketitle

Thermodynamics is indisputably a cornerstone of any description of physical processes occurring in macroscopic systems. 
Owing to its untold number of degrees of freedom, the average (bulk) behavior of a macroscopic system depicts a representative description of such complex systems, since fluctuations affecting its behavior are heavily suppressed. 
Such a feature provides unmatched robustness and generality for its predictions, allowing for a neat and minimalist picture of physical processes occurring in those systems \cite{Callen}.

However, with the advent of quantum technologies and the improvement of means for assessing biological machines, it became unavoidable having to deal with conditions for which thermodynamics loses accuracy: out-of-equilibrium small scale systems. 
Even more, as the system gets smaller, quantum effects become relevant, introducing a new physical origin for fluctuations observed in their processes. 
In this context, it is a fair question whether the laws of thermodynamics are still valid \cite{XuerebNJP}.
This quest led to the development of stochastic thermodynamics \cite{SeifertRP} as well as of quantum thermodynamics, with results ranging from classical and quantum fluctuation theorems \cite{Jarzynski,HanggiPRE,TalknerNatPhys} to connections between thermodynamics and quantum information \cite{BrandaoNatPhys}.
Research on quantum thermal machines \cite{XuerebNJP} fomented particular interest in 
the interconversion principles between heat and work \cite{lenard78} in cyclic processes \cite{BrandaoPNAS,goold15,cyril15}, in the quantum domain. 

In this paper, we put forward a realistic procedure for work extraction from one of the most elementary quantum physical engines, namely, a two-level system (TLS) and a spontaneously emitted propagating single photon.
Through a cyclic evolution involving a single photon packet (quantum piston) and a TLS state (working substance), we show that there can exist work extraction from a completely passive state \cite{lenard78}, which remains passive throughout the entire cycle.
Therefore, our proposal is complementary to the idea of extracting work from a quantum battery, where no cycle is performed and the initial quantum battery state must be an activated state (non-passive state) \cite{GooldNJP}.


{\it Work and Heat.-}
Average work $W_{\mathrm{net}}$ performed on the TLS and average heat $Q_{\mathrm{net}}$ absorbed by the TLS are defined in a general framework according to energy conservation \cite{alicki79},
\beq
\langle \dot{E} \rangle =  \dot{W} + \dot{Q},
\eeq
where $\langle E \rangle = \mbox{Tr}\{\rho_s(t) H_s(t)\}$,
$\dot{W} = \mbox{Tr}\{ \rho_s(t) \dot{H}_s(t)\} $ and 
$\dot{Q} = \mbox{Tr}\{ \dot{\rho}_s(t) H_s(t)\} $.
Here, $\rho_s(t)$ is the reduced density operator of the TLS.
$H_s(t)$ is the effective TLS Hamiltonian, already taking field effects into account. 
This is derived in Eq.(\ref{ME}).
No work is performed on the TLS if no change occurs in the TLS Hamiltonian, 
$\dot{W} = 0$ if $\dot{H}_s(t) = 0$.
Because only nonunitary evolution of the TLS contributes to the term $\dot{Q}$,
it can be regarded as heat absorption by the TLS from the environment.
The net work reads $W_{\mathrm{net}} = \int_{t_i}^{t_f} \dot{W}(t)dt$
and the net heat, $Q_{\mathrm{net}} = \int_{t_i}^{t_f} \dot{Q}(t)dt$, computed
from time $t_i$ up to $t_f$.
This definition of work is useful to define the concept of passivity.
A passive state is that from which no work can be extracted by means of a cyclic unitary protocol \cite{lenard78},
i.e., by a time-dependent Hamiltonian, from time $t_i$ to $t_f$, so that $H_s(t_f) = H_s(t_i)$.
It has been shown \cite{lenard78} that states that are diagonal in the energy basis and present no population inversion 
are passive.
Some states that are passive for a single copy of the state may not be passive for multiple copies of the state \cite{acin15}.
This is called activation.
States that cannot be activated regardless of the number of copies are called completely passive states.
Thermal equilibrium Gibbs states are an important class of completely passive states \cite{lenard78}.
Also, coherences or population inversion can break the passivity of a quantum state.

\begin{figure}[!htb]
\centering
\includegraphics[width=\linewidth]{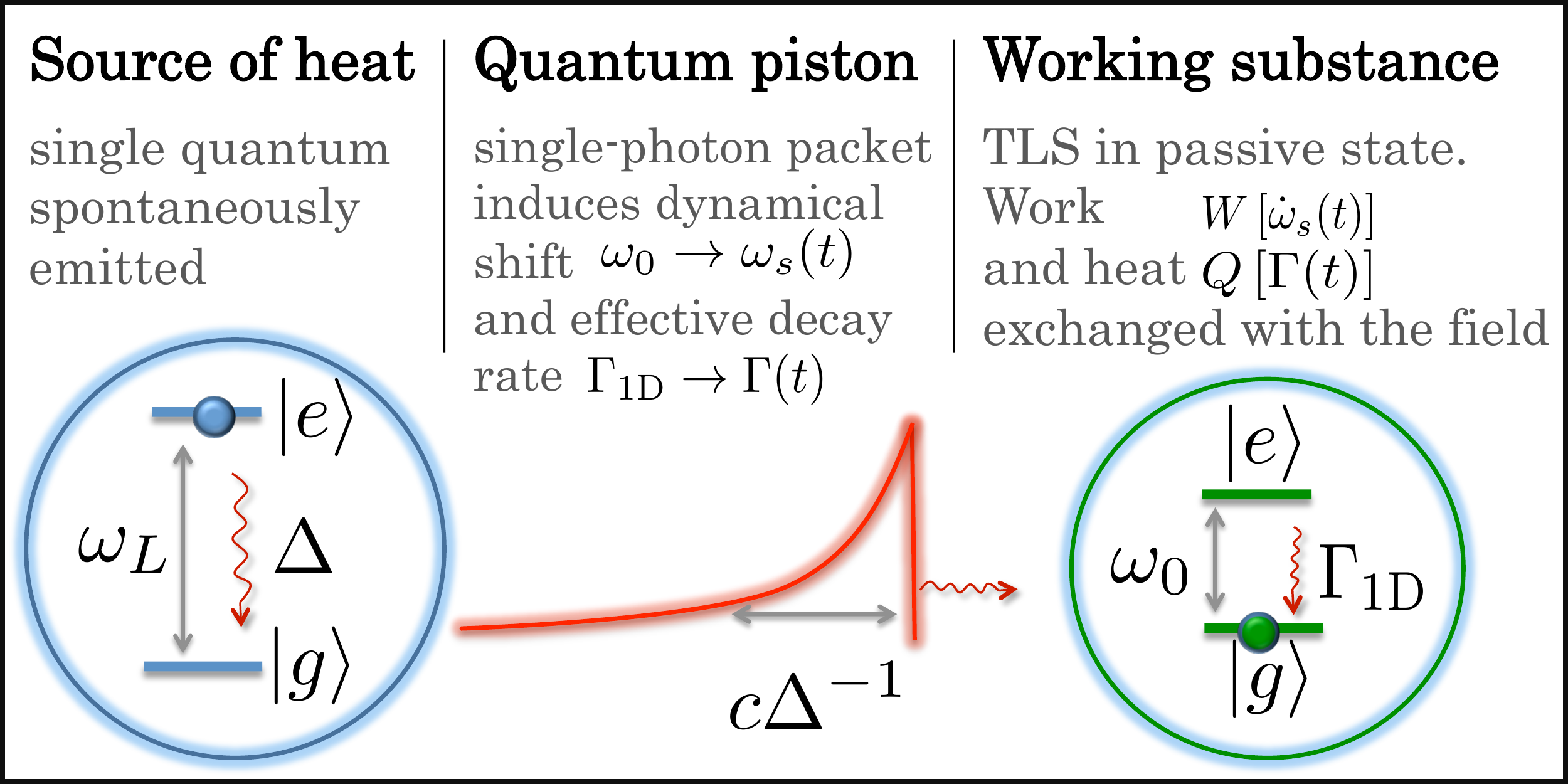}
\caption{
{\bf Elementary heat to work conversion.}
The photon spontaneously emitted in the form of heat both excites (heats up) and extracts work from an off-resonant two-level system (TLS), initially in a completely passive state $\ket{g}$. 
The photon packet modulates the TLS transition frequency in time, $\omega_s(t)$, playing the role of a quantum piston. 
The vacuum state of the field plays the role of a cold reservoir, inducing relaxation of the TLS towards $\ket{g}$. 
The process is cyclic in the sense that the initial and final states of the TLS are equal, 
$\ket{g} \rightarrow \ket{g}$.
} 
\label{model}
\end{figure}

In the present study scenario, the physical origin of both $\dot{W}$ and $\dot{Q}$
are due to a single-photon packet traveling through and interacting with the TLS.
Before the photon-packet reaches the TLS, the initial TLS state is assumed to be the ground state $\ket{g}$.
In that case, the frequency of the free TLS is $\omega_0$.
During the passage of the photon packet, a time-dependent modulation of the TLS transition frequency may be induced, 
$\omega_0 \rightarrow \omega_s(t)$, 
in which case the photon plays the role of a quantum piston (see Fig.\ref{model}).
Therefore, a finite amount of work may be extracted from the TLS, 
$\dot{\omega}_s(t) \neq 0\ \Rightarrow {W}_{\mathrm{net}} \neq 0$.
In turn, this will result in a time variation of the color of the reemitted photon.
That will be shown to happen if and only if the central frequency of the photon $\omega_L$ is not in resonance with the frequency of the free TLS $\omega_0$, that is, at nonzero detuning $\delta \equiv \omega_L - \omega_0 \neq 0$.
The quantum state corresponding to an excited TLS also corresponds to the field in the vacuum state,
which plays the role of a cold reservoir.
It induces relaxation of the TLS into the ground state $\ket{g}$
for times much longer than the typical pulse duration, $t\gg \Delta^{-1}$.
So, the process is cyclic, $\ket{g} \rightarrow \ket{g}$.
Therefore, the total energy difference is zero, $\langle E_{t \rightarrow \infty} \rangle - \langle E_{t=0} \rangle = 0$.
This establishes a relation between the total amount of work and heat on the TLS during a cycle, namely,
$W_{\mathrm{net}} = - Q_{\mathrm{net}}$.
%
This means that a net amount of work can only be extracted from the TLS, $W_{\mathrm{net}} < 0$, 
if the photon provides a positive amount of heat to the TLS, $Q_{\mathrm{net}} > 0$.
We are going to show below Eq.(\ref{ME}) that heat flux is a product
of the TLS excited state population $P_e(t)$, frequency $\omega_s(t)$ and
the negative of a time-dependent decay rate $\Gamma(t)$, i.e., 
$\dot{Q} = - \Gamma(t) \omega_s(t) P_e(t)$ (in $\hbar = 1$ units).
Whereas population and frequency are always positive, the effective decay rate may become negative.
Negativity of the effective decay rate at some time $t$ is a necessary condition for extracting a finite amount of work from the TLS, i.e.,
${W}_{\mathrm{net}} < 0\ \Rightarrow \ \Gamma(t) < 0$.
It has been shown \cite{vw16,hall14} that a negative decay rate is a signature of non Markovianity in the TLS dynamics.
It is also worth of attention that the quantum state of the TLS remains passive throughout the entire cycle of interaction with the photon packet, i.e., neither population inversion occurs, $P_e(t)<1/2$, nor coherences on the TLS reduced state are built during the TLS dynamics studied.

The analysis of the energy content of the photon packet shows that it is emitted in the form of pure heat.
The frequency of the emitter keeps time independent during the spontaneous emission.
This holds true even in the extreme case of an emitter in complete population inversion, starting at the excited state 
$\ket{e}$, for which the emitter initial state is no longer passive.
If the emitter is initially in a thermal Gibbs state at some finite temperature $T$, so that the initial excited state population is 
$p_{e}(T) \leq 1/2$, then the work extracted from the TLS is multiplied by the same factor $p_{e}(T)$ with respect to the case in which the emitter is completely inverted.

{\it Model.- }
A TLS coupled to a controllable electromagnetic environment well describes a wide
range of state-of-the-art experimental setups in quantum physics \cite{vuckovic1}.
In particular, one-dimensional (1D) electromagnetic environments offer the technical advantage
of allowing for broadband modes of propagating single-photon packets, whereas keeping strong matter-field interaction at the single-photon level \cite{domokos, fan05, chang07,danielNJP}. 
This kind of setup can provide single-photon sources \cite{lodahl08,jmg,inah13,wallraff},
single-photon packet spectroscopy \cite{sand12}, single-atom excitation with a single-photon packet \cite{ck} and single-atom transistors \cite{sand09,tsai10}, just to name a few examples.
Realistic 1D electromagnetic environments achieve guiding efficiencies from $95\%$ to almost $100\%$ \cite{jmg,tsai10}.


The dynamics of the composite system is unitary, governed by the total Hamiltonian 
$H = H_{\mathrm{system}}+H_{\mathrm{int}}+H_{\mathrm{field}}$.
Here,  
$H_{\mathrm{system}} = \hbar \omega_0 \sigma_{+}\sigma_{-}$
is the Hamiltonian of the free TLS, in which $\sigma_{-} = \sigma_{+}^\dagger = \ket{g}\bra{e}$.
The Hamiltonian of the 1D free field modes that propagate forwards $a_\omega$ and backwards $b_\omega$, with frequency $\omega$, inside the waveguide reads
$H_{\mathrm{field}} = \sum_{\omega} \hbar \omega [a^\dagger_\omega a_\omega + b^\dagger_\omega b_\omega] $.
The TLS-field interaction Hamiltonian, in the dipole and rotating-wave approximations, is given by \cite{domokos}
$H_{\mathrm{int}} = 
\sum_\omega -i \hbar g_\omega 
[\sigma_{+} (a_\omega e^{+i k_\omega x_s}+ b_\omega e^{-i k_\omega x_s}) - \mbox{H.c.}]$,
in which $x_s$ is the position of the TLS, 
$k_\omega = \omega/c$ is the wavevector modulus, $c$ is the speed of light, $\hbar g_\omega$ is the coupling energy
and H.c. denotes the Hermitian conjugate.
The reference frame is set so that $x_s = 0$.
We are interested in the zero- and one-excitation subspace, as described by the pure state
$\ket{\xi(t)} = c_0\ket{g,0} + \psi(t) \ket{e,0} +
\sum_\omega [\phi^{(a)}_\omega(t) a^\dagger_\omega + \phi^{(b)}_\omega(t) b^\dagger_\omega ] \ket{g,0}$,
where $\ket{0}$ is the vacuum state of the field.
The excited-state population of the TLS is $|\psi(t)|^2$.
The amplitude $c_0$ is time independent and we choose $c_0 = 0$.
The real-space representation of the field reads 
$\phi^{(a)}(x,t) = \sum_{\nu} \phi^{(a)}_{\nu}(t) e^{ik_{\nu}x}$ for the $a_\nu$ modes.
For the  $b_\nu$ modes, one substitutes $k_\nu$ for $-k_\nu$.
The photon packet prepared by means of a spontaneous emission has an exponential profile \cite{sand12},
$\phi^{(a)}(x,0) = N \Theta(-x) \exp[(\Delta/2 + i\omega_L)x/c]$, where 
$N = \sqrt{2\pi \rho_{\mathrm{1D}}\Delta}$
is a normalization factor and $\Theta(x)$ is the Heaviside step function.
$\rho_{\mathrm{1D}}$ is the density of states of the 1D continuum of frequencies, 
$\sum_\nu \rightarrow \int d\nu \rho_{\mathrm{1D}}$.
The packet linewidth is characterized by $\Delta$, hence the typical time duration of the pulse is $\Delta^{-1}$.
The central frequency of the packet is $\omega_L$, that corresponds to the transition frequency of the emitter.
The continuum of frequency modes imposes a vacuum-induced decay rate to the TLS, $\Gamma_{\mathrm{1D}} = 4\pi g_{\omega_0}^2 \rho_{\mathrm{1D}}$. 
We solve Schr\"odinger equation $i\hbar \partial_t \ket{\xi(t)} = H\ket{\xi(t)}$ to find the dynamics of the composite system, 
TLS plus field.
An explicit expression for the excited state amplitude is found, which guarantees that no population inversion is occurring, 
$|\psi(t)|^2 < 1/2$ at all times $t$.


The reduced dynamics of the TLS is obtained by tracing out the field variables, 
$\rho_s(t) = \mbox{Tr}_{\mathrm{field}}\{ \ket{\xi(t)} \bra{\xi(t)} \}$.
It can be shown \cite{breuer07} that $\rho_s(t)$ obeys the following master equation,
\beq
\partial_t \rho_s(t) = -i [H_s(t),\rho_s(t)] + \mathcal{L}_t \{ \rho_s(t) \},
\label{ME}
\eeq
where $H_s(t) = \omega_s(t) \sigma_+ \sigma_-$ and
\beq
\mathcal{L}_t \{ \rho_s(t) \} 
= 
\Gamma(t) \left( \sigma_- \rho_s(t) \sigma_+ - \frac{1}{2} \{  \sigma_+ \sigma_-, \rho_s(t) \}  \right),
\label{NU}\eeq
where $\{.,.\} $ is the anticommutator.
In Eqs.(\ref{ME}) and (\ref{NU}), the TLS time-dependent frequency induced by the single-photon packet is defined by 
$\omega_s(t) \equiv -\mbox{Im}[\dot{\psi}(t)/\psi(t)]$
and the time-dependent decay rate induced by the quantized field is defined by
$\Gamma(t) \equiv  -2\ \mbox{Re}[\dot{\psi}(t)/\psi(t)]$.
These expressions provide the analytical means by which one can obtain both the effective unitary and non-unitary parts of the TLS dynamics.
The former is driven by the effective TLS Hamiltonian $H_s(t)$ and the latter, by the effective Lindbladian $\mathcal{L}_t$.
Both terms have the same physical origin, namely, interaction of the TLS with the 1D continuum of modes of the electromagnetic field, in which a single-photon packet can be initially prepared.
In our case, we have that
$\dot{W} = \dot{\omega}_s(t) |\psi(t)|^2$.
This clarifies the importance of the time-dependent frequency modulation $\dot{\omega}_s(t) \neq 0$ 
induced by the photon for extraction of finite net work, 
$W_{\mathrm{net}} \neq 0$.
Analogously,
$\dot{Q} = \omega_s(t) \partial_t  |\psi(t)|^2 = -\Gamma(t) \omega_s(t) |\psi(t)|^2$,
as previously stated.
It shows the role played by the negativity of the time-dependent decay rate, 
$\Gamma(t) < 0$, in the heating process of the TLS,
 $\dot{Q} > 0$.
 
\begin{figure}[!htb]
\centering
\includegraphics[width=\linewidth]{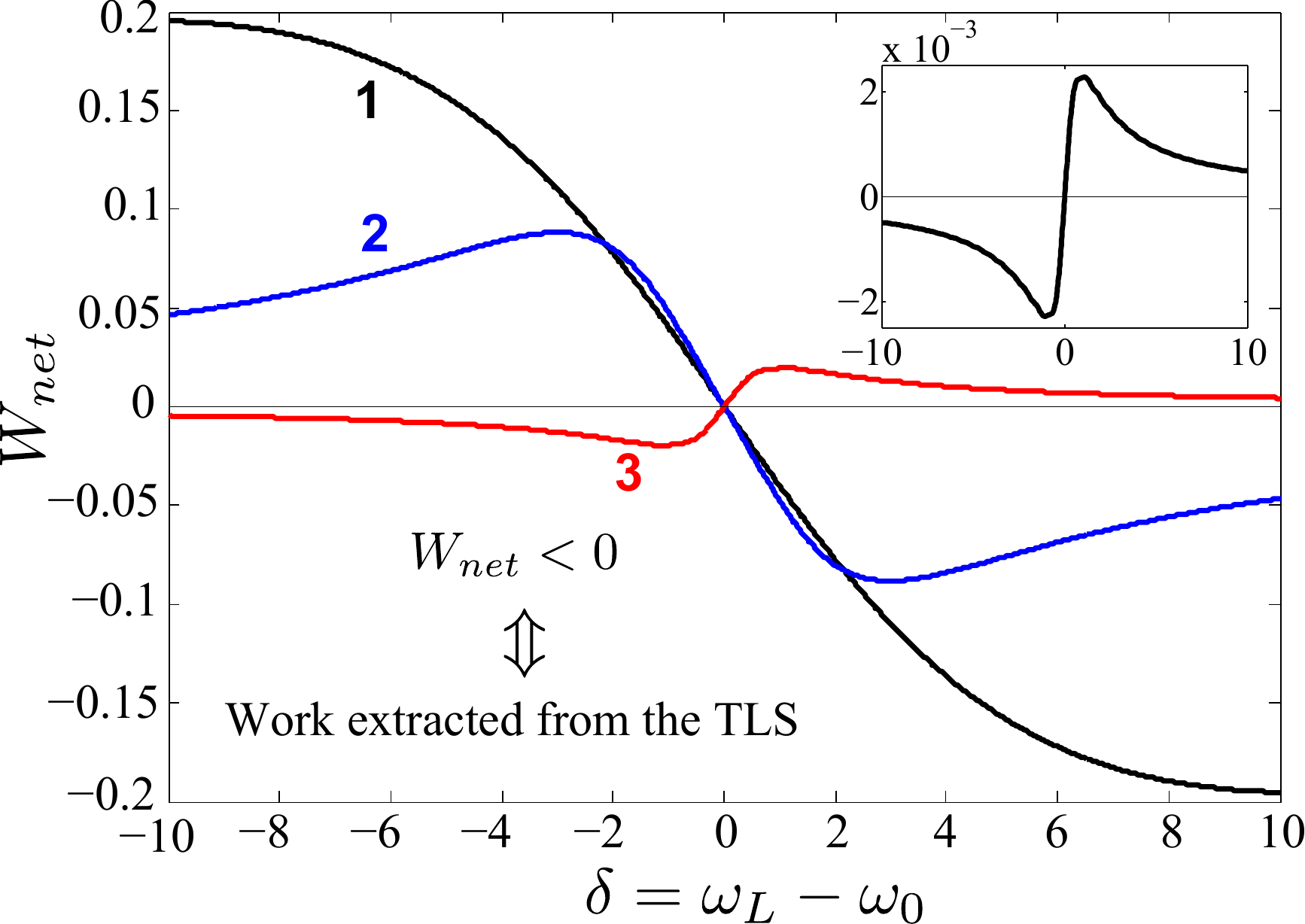}
\caption{
(Color online) {\bf Net work on the TLS.}
$W_{\mathrm{net}}<0$ means work performed by the TLS on the field.
The detuning is $\delta = \omega_L - \omega_0$.
The linewidths are 
$\Delta = 10$ (1-black), 
$2$ (2-blue), 
$0.1$ (3-red) and $0.01$ (inset).
$W_{\mathrm{net}}(\delta=0) = 0$, regardless of the packet linewidth.
To give a numerical reference, a work of $W_{\mathrm{net}} \sim 0.01$ ($\hbar = 1$) corresponds to the classical energetic cost of lifting a hydrogen atom under the gravity on earth by $\sim 1\AA$, that is nearly twice the Bohr radius.
} 
\label{w}
\end{figure}

The time-dependent TLS frequency evidenced by the master equation (\ref{ME}) has a sound physical meaning.
The propagation of the photon pulse causes disturbance in the electric field amplitude in time and space.
Since the TLS-field interaction $H_{\mathrm{int}}$ is given by means of a dipole coupling, a change in the electric field may stretch and contract the TLS dipole, creating a time-dependent Stark-shift effect.
To make this statement more precise, we can show that
$\dot{\omega}_s(t) = (1/2)\ \partial_t \left( \langle H_{\mathrm{int}} (t) \rangle / |\psi(t)|^2 \right)$,
where 
$\langle H_{\mathrm{int}} (t) \rangle = \bra{\xi(t)} H_{\mathrm{int}} \ket{\xi(t)}$.
So, the time variation of the TLS frequency is related to the time variation of the average dipole interaction energy, 
as intuitively expected.
In particular, if the average interaction energy is zero, the TLS frequency becomes static, $\dot{\omega}_s(t) = 0$.
We have found an expression in terms of the amplitudes of the photon packet and of the excite state,
$\langle H_{\mathrm{int}} (t) \rangle = 2g_{\omega_0} \mbox{Im}[\phi^{(a)}(0,t) \psi^*(t)]$.
That allows us to notice that, counterintuitively,  the average interaction energy 
$\langle H_{\mathrm{int}} (t) \rangle$ vanishes at all times at zero detuning $\delta = 0$.
An explicit expression for the time-dependent frequency has also been found,
$\omega_s(t) = 
\omega_0 + \partial_t \tan^{-1} \left( \sin(\delta t)/[\cos(\delta t) - \exp(\Delta - \Gamma_{\mathrm{1D}})t/2 ] \right) $.
Therefore, $\dot{\omega}_s(t)$ is an odd function of the detuning $\delta$.
Because $|\psi(t)|^2$ is an even function of $\delta$, we have that
$W_{\mathrm{net}}(-\delta) = - W_{\mathrm{net}}(\delta)$. 
Hence, the functionality of the TLS can be switched from consuming work  ($W_{\mathrm{net}}>0$) to delivering work ($W_{\mathrm{net}}<0$) by changing the sign of the detuning between the photon frequency and the bare TLS frequency.
Also, work vanishes at resonance, $W_{\mathrm{net}}(\delta = 0) = 0$.
In Fig.(\ref{w}), $W_{\mathrm{net}}$ is numerically computed as a function of $\delta$ for some linewidth values 
$\Delta = 0.01,\ 0.1,\ 2$ and $10$.
At arbitrarily large detunings, $\delta \gg \Gamma_{\mathrm{1D}}+\Delta$, the work vanishes, 
$W_{\mathrm{net}} \rightarrow 0$.
This is expectable, since an arbitrarily detuned TLS becomes transparent for the photon.
We notice that, given a finite detuning, the sign of the work depends on the sign of 
$ \Gamma_{\mathrm{1D}}-\Delta$.
Therefore, it is possible to identify a change in the behavior of the working substance depending
on the regime of the photon packet, i.e., the long packet ($\Delta^{-1} \gg \Gamma_{\mathrm{1D}}^{-1}$)
or the short packet ($\Delta^{-1} \ll \Gamma_{\mathrm{1D}}^{-1}$) regime.
Under the mode-matching condition, $\Delta = \Gamma_{\mathrm{1D}}$, work also vanishes.
It is particularly interesting to note that the work vanishes in the monochromatic (long packet) limit as well,
$W_{\mathrm{net}}(\Delta \ll \Gamma_{\mathrm{1D}}) \rightarrow 0$.
In that limit, the photon induces only a static frequency shift in the TLS,
$\omega_s(t) \rightarrow \omega_0 + \delta$, so that $\dot{\omega}_s(t) \rightarrow 0$,
which explains a vanishing net work. 
Since $\omega_s(t) \approx \omega_0 + \delta = \omega_L$, 
the TLS frequency becomes equal to the photon central frequency. 
Therefore, the outgoing photon has the same frequency as the incoming one.
 
The photon packet that both heats up and extracts work from the TLS is emitted in the form of pure heat.
This holds true even in the extreme case where the source atom, here called emitter, is initially in the excited state,
which is not passive due to population inversion.
To show that, we compute the time-dependent frequency of the emitter, $\omega^{\mathrm{SE}}_s(t)$.
If the initial state of the field is vacuum, spontaneous emission occurs, so that the excited state amplitude is
$\psi^{\mathrm{SE}}(t) = \exp-(\Gamma_{\mathrm{1D}}/2 + i\omega_L)t$.
Hence, $\omega^{\mathrm{SE}}_s(t) = -\mbox{Im}[\dot{\psi}^{\mathrm{SE}}/\psi^{\mathrm{SE}}] = \omega_L$.
Also, $\Gamma^{\mathrm{SE}}(t) = \Gamma_{\mathrm{1D}}$.
Because $\dot{\omega}^{\mathrm{SE}}_s = 0$, no work is performed by the emitter on the field.
The heat emitted, on the other hand, corresponds to the total amount of energy initially stored in the emitter,
$Q^{\mathrm{SE}}_{\mathrm{out}} = -Q^{\mathrm{SE}}_{\mathrm{in}} 
=- \int_0^\infty dt\ \omega_L \partial_t |\psi^{\mathrm{SE}}(t)|^2 = -\omega_L ( |\psi^{\mathrm{SE}}(\infty)|^2  -  |\psi^{\mathrm{SE}}(0)|^2)$. Therefore, the spontaneously emitted photon arises entirely in the form of heat, 
$Q^{\mathrm{SE}}_{\mathrm{out}} = \omega_L$.


{\it Field signatures of work and heat.-}
The output channels of the waveguide that establishes the 1D electromagnetic environment give information about the TLS dynamics. 
Hence, heat and work are both measurable in the outgoing electromagnetic field.
At channel $b$, for instance, the field amplitude $\phi^{(b)}(x_d,t)$ at the position of the detector $x_d < 0$ at time $t$ is directly proportional to the excited state amplitude, 
$\phi^{(b)}(x_d,t) = \sqrt{\Gamma_{\mathrm{1D}} \pi \rho_{\mathrm{1D}}} \psi(t-|x_d|/c)$.
The probability density of detecting a photon at position $x_d$ at time $t$, related to the intensity of the electric field at the detector, is 
$|\phi^{(b)}(x_d,t) |^2 \propto |\psi(t-|x_d|/c)|^2$.
Given that $\dot{Q} \propto \partial_t |\psi(t)|^2$, the time derivative of the probability of detecting a photon at time $t$ gives a signature of the heat that was being absorbed by the TLS at the previous time $t-|x_d|/c$.
Likewise, the signature of the time dependency of $\omega_s(t)$ also appears on the field $\phi^{(b)}(x_d,t)$.
By writing $\psi(t) = |\psi(t)|\exp{i\theta(t)}$, one promptly shows that $\omega_s(t) = -\dot{\theta}(t)$.
By the linearity of the field amplitude with respect to the excited-state amplitude, the phase of the field is given by $\theta_b(t) = \theta(t)$, so that $\dot{\theta}_b(t) = \dot{\theta}(t)$.
Because $\phi^{(b)}(x_d,t)$ represents the electric field amplitude, the negative of the time derivative of its phase, $-\dot{\theta}_b(t)$, can be intuitively interpreted as an effective time-dependent color of the field $\omega_{\mathrm{eff}}(t) \equiv -\dot{\theta}_b(t)$.
Therefore, $\omega_{\mathrm{eff}}(t) = \omega_s(t)$, i.e., the instantaneous color of the emitted field is equal to the instantaneous transition frequency of the emitter.
Indeed, $\omega_{\mathrm{eff}}(t) = \omega_L$ for the spontaneously emitted packet.
In addition, the temporal modulation of $\omega_{\mathrm{eff}}(t) \equiv \omega_0 + f(t)$ for the outgoing packet is very slow, $f(t) \sim \delta \ll \omega_0$.
These two properties corroborate the interpretation of $\omega_{\mathrm{eff}}(t)$ as a time-dependent color for the reemitted photon.
Besides, the effective time-dependent color described by $\omega_{\mathrm{eff}}(t)$ is more than a simple interpretation.
This quantity can be accessed experimentally, by means of a time-dependent fluorescence spectroscopy of the signal coming from channel $b$.
This spectrum is the Fourier transform of the two-time correlation function of the field, proportional to $\langle \sigma_+(t') \sigma_-(t) \rangle$.
For the spontaneously emitted photon, a Lorentzian peak is built around the central frequency $\omega_L$, in the spectrum.
For the reemitted photon, though, the peak position will vary in time during the buildup of the spectrum.
As a final consideration, it is important to state that energy conservation is guaranteed to the field after the TLS returns to its ground state.
The average energies $E_a = \sum_\omega \hbar \omega \langle a_\omega^\dagger a_{\omega} \rangle$ 
and $E_b = \sum_\omega \hbar \omega \langle b_\omega^\dagger b_{\omega} \rangle$,
in channels $a$ and $b$, satisfy the condition 
$E_a+E_b = \hbar \omega_L$.

{\it Conclusions.-}
We have shown that a single photon packet is capable of both heating and extracting work from a single TLS.
This effect is found for a photon as it is spontaneously emitted, therefore carrying only heat away from the emitter.
The TLS must be off resonance, $\omega_0 \neq \omega_L$, and off mode matching, 
$\Gamma_{\mathrm{1D}} \neq \Delta$, 
to allow for exchange of a finite amount of work $W$.
The TLS starts in a completely passive, remaining passive throughout the entire interaction time with the single-photon packet.
The physical meaning of the work performed by the TLS is found to be a dynamical change in the color of the photon during the reemission process. 
This can be measured in fluorescence spectroscopy experiments within state-of-the-art setups.
%
A very timely perspective opened by this paper is the question of whether it would be possible to convert the work extracted from a TLS into useful mechanical work in optomechanical devices. 
It would be particularly interesting to investigate that in systems allowing for single-photon packet propagation as, 
e.g., in Refs.\cite{inah13,cyril15}.


\begin{acknowledgements}
D.V. is supported by CNPq (Brazil) through Grant No. 477612/2013-0.
F.B. is supported by Instituto Nacional de Ci\^encia e Tecnologia -- Informa\c c\~ao Qu\^antica (INCT-IQ).
T.W. is supported by CNPq (Brazil) through Grant No. 478682/2013-1. 
\end{acknowledgements}


%

%

%

%

%
\end{document}